\documentclass{ws-procs9x6}

\begin{document}

\title{FROM EDELWEISS-I TO EDELWEISS-II}

\author{V. SANGLARD \\ for the EDELWEISS Collaboration}

\address{Institut de Physique Nucl\'eaire de Lyon, \\
Universit\'e Claude Bernard Lyon 1, \\
43, Bd du 11 Novembre 1918, \\ 
69622 Villeurbanne Cedex, FRANCE\\
E-mail: sanglard@ipnl.in2p3.fr}

\maketitle
\abstracts{
The EDELWEISS experiment is a Direct Dark Matter Search using 320~g heat-and-ionization
Ge cryogenic detectors. The final results obtained by the EDELWEISS-I stage 
corresponding to a total of 62~kg.day are presented. The status of EDELWEISS-II, 
involving in a first phase $\sim$~10~kg of detectors and aiming to gain two orders of magnitude in 
sensitivity, is also described.}
\section{Introduction}
The EDELWEISS experiment is dedicated to the search for non-baryonic cold dark matter in 
the form of WIMPs (Weakly Interactive Massive Particles). The direct detection principle 
consists in the measurement of the energy released by nuclear recoils produced in an ordinary matter target 
by elastic collisions of WIMPs from the galactic halo.\\
The EDELWEISS detectors are cryogenic Ge bolometers with simultaneous measurement of
phonon and ionization signals. The comparison of the two signals provides an excellent
event-by-event discrimination between nuclear recoils (induced by WIMP or neutron 
scattering) and electronic recoils (induced by $\beta$ or $\gamma$-radioactivity).\\
The experiment is located in the Modane Underground Laboratory in the tunnel 
connecting France and Italy under $\sim$1800~m of rock ($\sim$4800~mwe). 
In the laboratory, the resulting muon flux is 4~$\mu$/m$^2$/d and the fast neutron flux has been 
measured~\cite{neutron} to be $\sim$~1.6$\times$ 10$^{-6}$~cm$^2$/s.\\
During three years, 62~kg.day of data have been accumulated with five
320~g Ge detectors.
\section{Experimental set-up}
Between 2002 and 2003, three 320~g Ge detectors were operated 
simultaneously in a dilution cryostat
with a regulated temperature of 17~$\pm$~0.01~mK. A passive shielding made of paraffin (30~cm), 
lead~(15cm) and copper~(10~cm) surrounded the experiment~\cite{{edel1},{edel2}}.\\
The detectors~\cite{edel3} are made of a cylindrical Ge crystal with Al 
electrodes to collect ionization signals and a NTD heat sensor glued onto one electrode
to collect the phonon signal. The top electrode is segmented in a central electrode and 
an annular guard ring and defines a fiducial volume corresponding to 
57~$\pm$~2~$\%$ of the total volume~\cite{edel4}. On four of the five detectors used in 
EDELWEISS-I an amorphous layer (either of Ge or Si) was deposited under the electrodes to improve 
charge collection of near surface events~\cite{surface}. \\
The resolutions and energy thresholds of the detectors are summarized in Ref.~\cite{edel4}. 
Detectors with an amorphous layer show a 99.99~$\%$ gamma rejection at a recoil energy of 
100~keV and a 99.9~$\%$ gamma rejection at a recoil energy of 15~keV.
\section{Edelweiss-I results}
\begin{figure}[htbp]
\begin{center}
\includegraphics*[width=8cm,height=8cm]{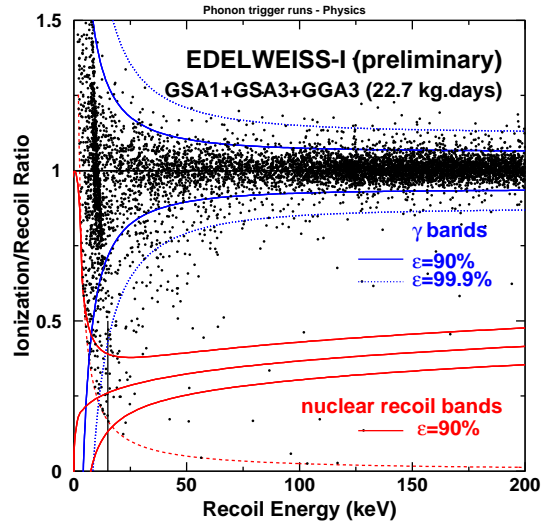}
\end{center}
\caption[]{$\frac{E_I}{E_R}$ versus $E_R$ (fiducial volume) for physics runs with the  
phonon trigger configuration.}
\label{data}
\end{figure}

\begin{figure}[htbp]
\begin{center}
\includegraphics*[width=8cm,height=8cm]{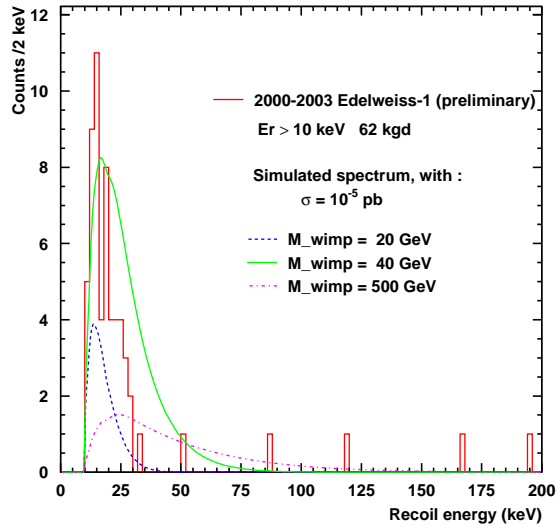}
\end{center}
\caption[]{Energy spectrum for the EDELWEISS-I data, for $E_R$~$>$~10~keV, compared to simulated 
theoretical WIMP spectrum  for M$_{WIMP}$ = 20, 40, 500~GeV/c$^2$.}
\label{spectrum}
\end{figure}

\begin{figure}[htbp]
\begin{center}
\includegraphics*[width=8cm,height=8cm]{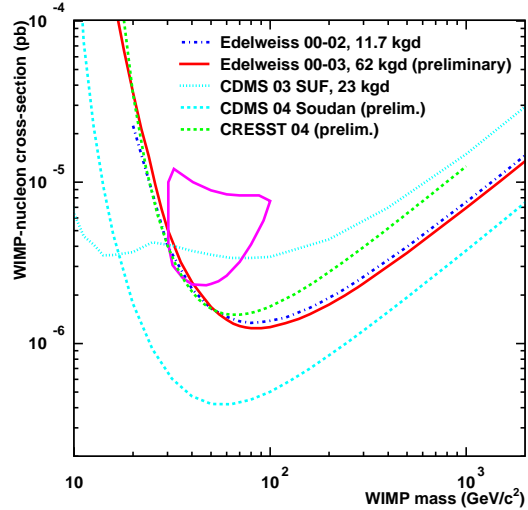}
\end{center}
\caption[]{Preliminary spin independent exclusion limit for EDELWEISS-I without 
background subtraction compared to CDMS limit, when the experiment was operated
 at Stanford~\cite{cdms1} and in the Soudan mine~\cite{cdms2}, and the CRESST 
 limit~\cite{cresst}. The closed contour is the allowed region at 3$\sigma$ C.L. from 
 DAMA NaI1-4 annual modulation data~\cite{dama1}.}
\label{exclusion}
\end{figure}

During the EDELWEISS-I stage (2000-2003), four physics runs have been performed with five 
detectors. In the three first runs, the trigger was the fast ionization signal. For the
last run, the trigger was the phonon signal. Thanks to a better resolution and the absence of 
quenching factor on the phonon signal, the phonon trigger improves the efficiency 
at low energy for nuclear recoils. In this last configuration, a $\sim$~100~$\%$ efficiency 
has been reached at 15~keV on the three detectors. With the ionization trigger a $\sim$~100~$\%$ 
efficiency was reached at 20~keV or 30~keV, depending on the detector. 
The low-background physics data recorded in the phonon trigger configuration are shown 
in Fig.~\ref{data}.\\
Considering the complete 62~kg.day data set, 60 events compatible with nuclear recoils have 
been recorded above a recoil energy of 10~keV. The corresponding 
energy spectrum is shown in Fig.~\ref{spectrum}, compared with simulations of theoretical 
spectrum for different WIMP masses, taking into account the recoil energy dependence 
efficiency\footnote{The efficiency calculation takes into account thresholds, resolutions, the 
90~$\%$ efficiency (1.65~$\sigma$) for the nuclear recoil band and the 99.9~$\%$ rejection 
(3.29~$\sigma$) of the electronic recoils.} of all experimental configurations. The 
overall shape of the experimental spectrum is incompatible with WIMP masses above 20~GeV/c$^2$.\\
Setting our analysis threshold to 20~keV, above which the experimental efficiency is greater than
75~$\%$, 
23 events compatible with nuclear recoils have been observed. Considering all these events 
as possible WIMP interactions and taking into account the efficiency versus recoil
energy function of each run\footnote{For example, for the complete data set of Edelweiss-I a 
50~$\%$ efficiency is reached for a recoil energy of 15~keV.}, a conservative upper limit 
on the WIMP-nucleon cross-section as a function of the WIMP mass has been derived with the 
Optimum Interval Method~\cite{yellin}. This method allows to compute an exclusion limit in the 
presence of an unknown background. Fig.~\ref{exclusion} shows the EDELWEISS-I spin independent 
exclusion limit, assuming a standard spherical and isothermal galactic WIMP halo with a local 
density of 0.3~GeV/c$^2$/cm$^3$, a rms velocity of 270~km/s, an escape velocity of 650~km/s 
and a relative Earth-halo velocity of 230~km/s. Limits from other running experiments are also 
shown on Fig.~\ref{exclusion}. With no background subtraction and an extended exposure, the 
new limit is consistent with the previous published one~\cite{edel}.\\
Although the EDELWEISS-I limit of Fig.~\ref{exclusion} is derived assuming all events as 
possible WIMP candidates, the experimental data reveal some clues as to the nature of 
possible backgrounds. For example, a two detector coincidence 
between nuclear recoils has been recorded. This event is very likely a neutron-neutron 
coincidence, indicating that a certain fraction of events in Fig.~\ref{spectrum} could be due
to single hits by neutrons. Miscollected charge events, as indicated by the few events lying 
between electronic and nuclear recoil bands in Fig.~\ref{data}, are another possible source of 
background, because they can simulate nuclear recoils. But with the present statistics, 
limited largely by the number of detectors (the EDELWEISS-I cryostat could not receive more 
than 3$\times$320~g detectors), it is not possible to conclude any further. 
\section{Lessons for EDELWEISS-II} 
The second phase of the experiment will be EDELWEISS-II, with an expected sensitivity of 
0.002~evt/kg/d. Specific improvements are aimed at reducing the possible background sources, 
that may have limited the sensitivity of EDELWEISS-I. In addition, the detector number will be 
increased up to 28 to achieve a Ge mass of $\sim$~10~kg in a first stage.
\subsection{Neutrons}  
The low energy neutron background, due to the radioactive surrounding rock, is attenuated 
by more than three orders of magnitude thanks to a 50~cm polyethylene shielding. 
In addition, a muon veto~\cite{veto} surrounding the experiment will tag muons interacting
in the lead shielding. The increased number of detectors will improve the possibility of 
detecting multiple interactions of neutrons. 
\subsection{Surface events}
Surface events, namely interactions near electrodes, show a deficit of the charge collection. 
One of the R$\&$D goals in EDELWEISS is the event-by-event identification of these miscollected 
events and their active rejection. 
A new generation of detectors has been developped with NbSi thin film sensors (instead 
of the NTD heat sensors for present detectors). They consist in a Ge crystal with 
two NbSi sensors acting also as electrodes for charge collection. These thin film sensors 
are sensitive to the athermal component of the phonon signal, acting as near-surface 
interaction tag~\cite{nbsi}. Several tests have been made in the EDELWEISS-I setup with 
three 200~g Ge detectors showing a reduction by a factor 10 of the surface event rate 
while retaining a 50~$\%$ efficiency. Seven 400~g Ge detectors are being prepared 
in a first stage for EDELWEISS-II.\\
Furthermore, improved radiopurity and clean room conditions are expected to reduce the contaminations
and the rate of surface electrons.
\section{Conclusion}
EDELWEISS-I experiment has reached its limit sensitivity near 10$^{-6}$~pb, allowing 
the exclusion of some optimistic SUSY models. The goal for the future with EDELWEISS-II  
is to reach more favored models close to 10$^{-8}$~pb.\\
The EDELWEISS-I experiment was stopped in March 2004 to allow the installation of the second 
phase EDELWEISS-II. The first runs will be performed with 21~$\times$~320~g Ge 
detectors with NTD heat sensor and 7~$\times$~400~g Ge detectors with NbSi thin 
film sensor. Data taking in the new setup is scheduled for end-2005.

\end{document}